# On RT CW THz Cyclotron Resonance lasing in graphene in crossed *E, H* fields

A. Andronov and V. Pozdniakova

In framework of classical consideration of electron trajectories in crossed *E, H* fields and conductivity of electron system on cyclotron resonance in single layer graphene possibility to achieve THz cyclotron lasing in hexagonal boron nitride–single layer graphene sandwiches is discussed. By simplified consideration with known data on scattering rate in the sandwiches it is demonstrated that the CW laser action can be achieved in high quality sandwiches at room temperature at frequencies above 1 THz in magnetic field above 10000 Gauss.

*Introduction:* Possibility of CR (cyclotron resonance) amplification in crossed E, H fields in a semiconductor due to electron accumulation at closed trajectories at electron energy below optical phonon energy where only low elastic scattering occurs due to low lattice temperature was first mentioned in [1]. CR THz lasing based on such mechanism was first demonstrated on light holes in p-Ge in [2]. Then a lot of group in the World extended that work (see e.g. [3, 4]). Though the lasers demonstrated very high (two-three fold) THz frequency tuning by change in E, H fields they were not broadly used in application because of liquid helium cooling and pulsed mode performance.

In recent years high quality high mobility graphene layers in hexagonal boron nitride–graphene sandwiches have been demonstrated. In particular THz amplification at 300K in gated sandwiches with single graphene layer (SGL) was reported [5] and THz detectors on gate plasmons in sandwiches with double graphene layers were demonstrated [6]. This February THz detector based on plasmons near CR second harmonic in similar sandwiches with SGL was demonstrated [7]. We have explained [8] results of THz amplification in [5] basing on THz negative conductivity of hot electrons in the sandwiches due to the momentum space electrons bunching at transit time resonance under streaming. Such effect was also discussed in [1] and demonstrated in [9]. These results demonstrate that in graphene there is a real possibility to get CW CR lasing in crossed E, H, fields at high temperature because in sandwiches optical phonon energy is about 2000K. And in this report possibility of CR amplification and lasing in SGL graphene in crossed E, H fields are discussed. This topic was initiated for us by Erich Gornik (Vienna) after considering our work [8]. As in p-Ge in graphene strong spontaneous optical phonon emission and non-equidistance of Landau levels should be responsible for Landau level population inversion and CR negative conductivity in crossed fields.

*Electron trajectories and THz CR conductivity in SGL in crossed fields:* In this, motivation, work we perform discussion of those problems within classical approach both for considering electron trajectories and CR conductivity calculation. Figure 1 gives electron trajectories and their CR frequencies in SGL in crossed fields with ("Dirac") electronic band $\varepsilon = V_F p$, $V_F$ – «Fermi» velocity, $p$ – electron momentum with components $p_x$, $p_y$, $E = E_x$ and $H = H_z$ in dimension less variables $r = p/p_0$, $\tau = \omega_c t$, $\omega_c = eH/mc$, $m = p_0/V_F$, $\varepsilon_0 = V_F p_0 = \hbar\omega_0$ – optical phonon energy, and at ratio $D = cE/HV_F = 0.5$. Also estimated electron distribution $f(\rho)$ integrated over full trajectory is shown ($\rho$ is radius of a trajectory approximated by circles). In region of inversion – $df(\rho)/d\rho$ approximate expression is: $f(\rho) \approx A\rho$ with $A = 3$ if we ignore contribution of trajectories entering region $\rho > 1$ ($\varepsilon > \varepsilon_0$). We will not go here in details of the distribution estimate. We say only that they refer to the case of CR frequency $\omega_c \gg \nu_{o0}$, $\nu_{o0}$ is characteristic optical phonon scattering rate. At such condition occupation of electron trajectories entering energy $\varepsilon > \varepsilon_0$ can be consider constant and approximately is equal. And distribution function on trajectories below an optical phonon energy $\varepsilon_0$ is determined by value of surface area at $p_x$, $p_y$ plane corresponding to specific trajectory (two example are shown in Figure 1) from which total filling of that trajectory by optical phonon emission occurs. The areas are situated between optical phonon energy and upper boundary of such transition to a trajectory shown. The areas are determined by change of electron momentum just by $p_0$ because due to Dirac electron band optical emission of optical phonon changes the momentum for $p_0$ (or $r$ for 1). At the same time ratio of incoming rate by optical phonon emission to outgoing rate by acoustical phonon scattering is independent from initial position of electron in the areas (due to specific density of states and the above feature). This demonstrates that only size of the areas is determined ratio of trajectory distribution functions. From Figure 1 it is clear that the area for trajectory with large $\rho$ is substantially high that for one with smaller $\rho$ that provide population inversion in distribution over $\rho$ – $df(\rho)/d\rho$. Presumably this estimate for distribution function stand also for $\omega_c \approx \nu_o$. It is this inversion which provides CR negative conductivity.

Under classical calculation under τ – approximation real part of CR conductivity in circular homogeneous resonant to electron rotation AC high frequency field for distribution function $f(\rho)$ is:

$$\operatorname{Re}\sigma_c(\omega) = -\frac{e^2 N_S}{2} \times \int \frac{df}{dp}\cdot\frac{p^2}{m}\cdot\frac{\upsilon(p)}{(\omega-\omega_c(p))^2-\upsilon^2(p)}\,dp \qquad (1)$$

Here $\upsilon(p) = 1/\tau$ scattering rate, $N_S$ – electron surface density and $\int pf dp = 1$. In limit $\upsilon(p) \to 0$

$$\frac{\upsilon(p)}{(\omega-\omega_c(p))^2-\upsilon^2(p)} \to \pi\delta(\omega-\omega_c(p)) \qquad (2)$$

$\delta(\omega - \omega_c(p))$ – delta-function. Now we have

$$\operatorname{Re}\sigma_c(\omega) = -\frac{e^2 N_S p^2}{2m}\cdot\frac{\pi}{|d\omega_c(p)/dp|}\cdot\frac{df}{dp} \qquad (3)$$

with $p = p^*$ found from $\omega = \omega(p^*)$. We see that as was said before if $(df/dp)_{p^*} > 0$ then $Re\,\sigma_c(\omega) < 0$.

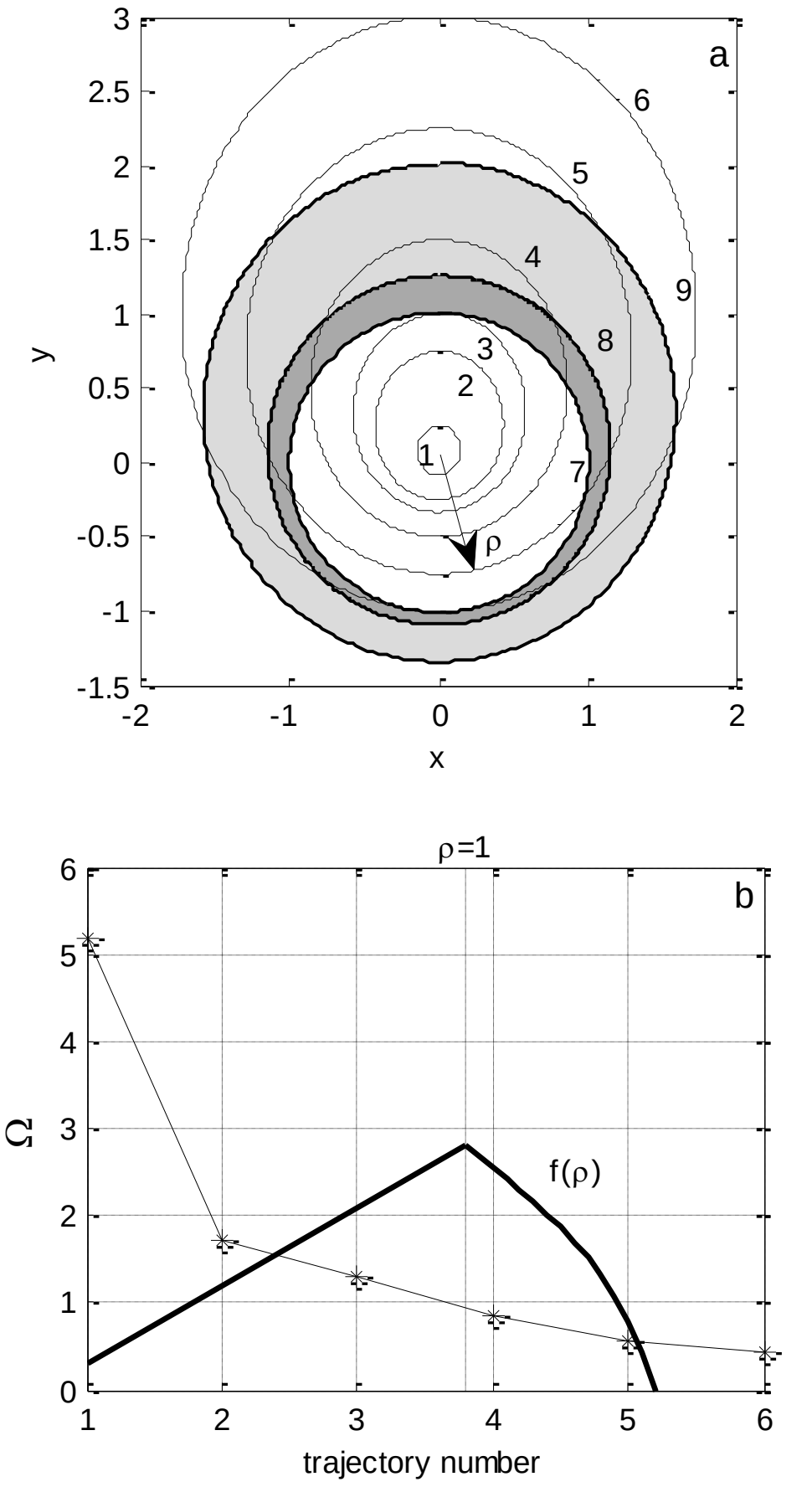

**Fig. 1** *(a) electron trajectories in SLG in crosses E, H fields for $D = cE/HV_F = 0.5$: shown trajectories (number 1 – 6), optical phonon energy (7) and upper boundaries for optical phonon emission at trajectory 1 (8) and 3 (9); areas involved in optical phonon emission are between optical phonon energy and the upper bounds indicated; ρ are trajectories radius. (b) CR frequencies on trajectories and distribution function on the trajectories versus their radius ρ.*

Conditions for expressions (2) and (3) with $\operatorname{Re}\sigma_c(\omega) < 0$ to be justified it is necessary scattering rate to be below CR frequency spread in the population inversion region. From data in Figure 1 on the spread at $\rho \approx 0.75$ (curve 3) we gat: $\upsilon(p) < 0.5\omega_c$.

To estimate $\mathrm{Re}\,\sigma_c(\omega) < 0$ value and CR frequencies and magnetic fields we use numbers from work [8]. They are: acoustical phonon rate $\upsilon_a(p) = \upsilon_{a0} p/p_0 = \upsilon_{a0}\rho$ and optical phonon rate (spontaneous emission at $\varepsilon > \varepsilon_0$) $\upsilon_0(p) = \upsilon_{oo}(p - p_0)/p_0 = \upsilon_{oo}(\rho - 1)$. Numbers in [7] are: $\upsilon_{a0} = 3 \cdot 10^{12}\mathrm{sec}^{-1}$ for $T$ = 300 K and $\upsilon_{oo} = 3 \cdot 10^{13}\mathrm{sec}^{-1}$.

Value of $\upsilon_{a0}$ used in [8] is well explaining electric field threshold value for streaming beginning in [5]. But it is twice lower than needed for mobility value (50 000 cm$^2$/(V·sec)) usually claimed for good sandwiched. Difference is attributed to impurity or disorder scatterings. Presumably in work [5] these scattering was not very important. Anyways we will use the above numbers for estimate conditions for the CR lasing supposing that it is possible to fabricate the sandwiches without substantial impurities or disorder. If they are important then region with $\mathrm{Re}\,\sigma_c(\omega) < 0$ only shift to higher THz frequencies and magnetic fields. It is important also that with low impurity or disorder scattering acoustical scattering rate is decrease at low temperatures broadening frequency region for the lasing. But we present here only estimates for $T = 300$ K. In the main condition for the CR negative conductivity to occur $\upsilon(p) < 0.5\omega_c$ for $\upsilon(p)$ we must chose averaged at over $\rho < 1$ value of $\upsilon_a(p) = \upsilon_{a0}\rho$ to get $\upsilon_{a0}/2$.

Now condition for the lasing is: $\upsilon_{a0} < \omega_c$ or $\omega_c > 3 \cdot 10^{12}\mathrm{sec}^{-1}$. To make condition more strong we double $\omega_c$ to have $\omega_c > 6 \cdot 10^{12}\mathrm{sec}^{-1}$ or for "simple" frequency $f_0 > 1$ THz. With $m = p_0/V_F \approx 0.03 m_0$ and for $\rho \approx 0.7$ chosen $\omega_c \approx 6 \cdot 10^{12}\mathrm{sec}^{-1}$ ($f_0 \approx 1$ THz) for $H = 10\,000$ while from (3) we have

$$|\mathrm{Re}\,\sigma_c(\omega)| \approx \frac{5e^2 N_S}{m\omega_c(p^*)} \qquad (4)$$

To chose $N_S$ e-e scattering must be taken into account: e-e rate to be below acoustical phonon rate. In work [10] e-e rate in SLG was calculated in discussion of high electric field transport at $T$ = 300 K. Average calculated e-e rate at energy below optical energy is $3 \cdot 10^{13}\mathrm{sec}^{-1}$ for $N_S = 7.7 \cdot 10^{12}\mathrm{cm}^{-2}$. The rate is proportional to $N_S$. And to have e-e rate below average acoustical phonon scattering rate ($1.5 \cdot 10^{12}\mathrm{sec}^{-1}$) it is necessary to divide that $N_S$ by 30 to get $N_S = 2.5 \cdot 10^{11}\mathrm{cm}^{-2}$. But the calculation in [10] refers to single graphene valley (because inter valley e-e scattering is negligible). Graphene Brillouin zone has 4 valleys (two of different spin and 6 one third of a valley at zone boundaries). So the overall $N_S$ in all valleys (what contribute to conductivity) should be $N_S = 10^{12}\mathrm{cm}^{-2}$.

Estimated value of $|\mathrm{Re}\,\sigma_c(\omega)|$ from (4) with $N_S = 10^{12}\mathrm{cm}^{-2}$ for $f_0 = 1$ THz ($\omega_c(p^*) = 6 \cdot 10^{12}\mathrm{sec}^{-1}$) is: $|\mathrm{Re}\,\sigma_c(\omega)| \approx 5 \cdot 10^9$ cm/sec. This is surface conductivity providing surface current $j_S$ in electric field $\tilde{E}$ of electromagnetic wave crossing this surface: $j_S = \mathrm{Re}\,\sigma_c(\omega)\,\tilde{E}$. This current in turn provide jump of the wave magnetic field $\tilde{H} = (4\pi/c)j_S = (4\pi/c)\,\mathrm{Re}\,\sigma_c(\omega)\,\tilde{E} = 2\tilde{E}$. So we got ratio $\tilde{E}/\tilde{H} = 1/2$ at the surface demonstrating high enough effect of surface current on electromagnetic wave interacting with the current showing possibility to use the CR negative conductivity in a laser. This value of the current effect we can compare with value on modulation at THz transmission cross similar sandwich without DC magnetic field measured in [5] – up to 9%.

*Conclusion:* We demonstrate by simple consideration that sandwich with SGL can serve as medium for CR THz laser in crossed *E, H* fields at $T$ = 300 K for THz frequencies higher than 1 THz. No doubt that the laser can work CW because due to magnetic field presented heating problem must be less important than that in work [1] where CW THz emission was demonstrated in high electric field without magnetic field at $T$ = 300 K. This situation is also due to the feature that at $\nu_{oo}/\upsilon_{a0} \approx 10$ main part of electrons is at closed trajectories at $\varepsilon < \varepsilon_0$ where scattering rate is low and main electric field is Hall field while dissipative part of electric field is low.

*Acknowledgments:* The authors are indebted to Erich Gornik (Vienna) who initiated this work. Work is performed as part of the State contract at IPM RAS № 0030-2021-0020.

A. Andronov and V. Pozdniakova (*Institute for Physics of Microstructures, Russian Academy of Sciences, Nizhny Novgorod, Russia*)

E-mail: andron@ipmras.ru

**References**

1. Andronov, A.A., Kozlov, V.A.: 'Low-temperature Negative Differential Microwave Conductivity in Semiconductors Following Elastic Scattering of Electrons', *Pisma v ZETF*, 1973, **17,** pp. 124–127
2. Ivanov, Yu.L.,Vasil'ev, Yu.B.: 'Submillimeter emission of hot holes in Ge in crossed magnetic field', *Pisma v Zhurnal Tekhnicheskoi Fiziki*, 1983, **9:10**, pp. 613–616
3. Gornik, E., Andronov, A.A. (eds): 'Far-infrared semiconductor lasers', *Optical and Quantum Electronics*, 1991, **23**, no. 1-2, doi: https://doi.org/10.1007/BF00619760
4. Pfeffer, P. et al.: 'p-type Ge cyclotron-resonance laser: Theory and experiment', *Phys. Rev. B*, 1993, **47**, no. 8, pp. 4522–4531, doi: https://doi.org/10.1103/PhysRevB.47.4522
5. Boubanga-Tombet, S. et al.: 'Room-Temperature Amplification of Terahertz Radiation by Grating-Gate Graphene Structures', *Phys. Rev. X*, 2020, **10**, pp. 031004-1–031004-19, doi: https://doi.org/10.1103/PhysRevX.10.031004
6. Bandurin, D.A., Svintsov, D., Gayduchenko, I. et al.: 'Resonant terahertz detection using graphene plasmons', *Nat. Commun.*, 2018, **9**, p. 5392, doi: https://doi.org/10.1038/s41467-018-07848-w
7. Bandurin, D.A., Mönch, E., Kapralov, K. *et al.:* 'Cyclotron resonance overtones and near-field magnetoabsorption via terahertz Bernstein modes in graphene.' *Nat. Phys.*, 2022, doi: https://doi.org/10.1038/s41567-021-01494-8
8. Andronov, A.A., Pozdniakova, V.I.: 'Terahertz Dispersion and Amplification under Electron Streaming in Graphene at 300 K', *Semiconductors*, 2020, **54**, no. 9, pp. 1078–1075, doi: https://doi.org/10.1134/S106378262009002X
9. Vorob'ev, L.E., Danilov, S. N., Tulupenko, V. N., Firsov, D. A.: 'Generation of millimeter radiation due to electric field induces transit time resonance in InP', *Journal of Experimental and Theoretical Physics Letters*, 2001, **73**, pp. 219–222, doi: https://doi.org/10.1134/1.1371057
10. Fang, T., Komar, A., Xing, H., Jena, H.: 'High Field Transport in Two-Dimensional Graphene', *Phys. Rev. B*, 2011, **84**, pp. 125450-1-125450-7, doi: https://doi.org/10.1103/PhysRevB.84.125450